\documentclass{article}
\usepackage[english]{babel}
\usepackage[a4paper]{geometry}
\usepackage{amsmath}
\usepackage{physics}
\usepackage{graphicx}
\usepackage[colorlinks=true, allcolors=blue]{hyperref}
\usepackage{stackengine}
\usepackage[super,sort&compress,comma]{natbib}
\usepackage{caption}
\usepackage{subcaption}
\usepackage{mwe}
\usepackage{soul,xcolor}
\usepackage{authblk}

\title{Forming and Compliance-free Operation of Low-energy, Fast-switching HfO$_x$S$_y$/HfS$_2$ Memristors}

\author[1,2]{Aferdita Xhameni}
\author[1,3]{AbdulAziz AlMutairi}
\author[4]{Xuyun Guo}
\author[3]{Irina Chircă}
\author[2]{Tianyi Wen}
\author[3]{Stephan Hofmann}
\author[4]{Valeria Nicolosi}
\author[1,2*]{Antonio Lombardo}

\affil[1]{London Centre for Nanotechnology, 19 Gordon St, London, WC1H 0AH, United Kingdom}
\affil[2]{Department of Electronic \& Electrical Engineering, Malet Place, University College London, WC1E 7JE, United Kingdom}
\affil[3]{Department of Engineering, University of Cambridge, Cambridge CB3 0FA, United Kingdom}
\affil[4]{School of Chemistry, Centre for Research on Adaptive Nanostructures and Nanodevices (CRANN) \& Advanced Materials Bio-Engineering Research Centre (AMBER), Trinity College Dublin, Dublin 2, Ireland}
\affil[*]{Corresponding author: Antonio Lombardo (email: a.lombardo@ucl.ac.uk)}

\date{}

\begin{document}
\maketitle

\begin{abstract}
%\textbf{}
\noindent

We demonstrate low energy, forming and compliance-free operation of a resistive memory obtained by the partial oxidation of a two-dimensional layered van-der-Waals semiconductor: hafnium disulfide (HfS$_2$). Semiconductor - oxide heterostructures are achieved by low temperature ($<300^{o}$C) thermal oxidation of HfS$_2$ in dry conditions, carefully controlling process parameters. The resulting HfO$_x$S$_y$/HfS$_2$ heterostructures are integrated between metal contacts, forming vertical crossbar devices. Forming-free, compliance-free resistive switching between non-volatile states is demonstrated by applying voltage pulses and measuring the current response in time. We show non-volatile memory operation with an R$_{ON}$/ R$_{OFF}$ of 102, programmable by 80ns WRITE and ERASE operations. Multiple stable resistance states are achieved by modulating pulse width and amplitude, down to 60ns, $<$ 20pJ operation. This demonstrates the capability of these devices for low - energy, fast-switching and multi-state programming. Resistance states were retained without fail at 150$^o$C over 10$^4$s, showcasing the potential of these devices for long retention times and resilience to ageing. Low-energy resistive switching measurements were repeated in vacuum (8.6 mbar) showing unchanged characteristics and no dependence of the device on surrounding oxygen or water vapour. Using a technology computer-aided design (TCAD) tool, we explore the role of the semiconductor layer in tuning the device conductance and driving gradual resistive switching in 2D HfO$_x$ - based devices.

\end{abstract}

\textbf{Keywords}: \textit{resistive switching (RS), resistive memories, memristors, neuromorphic computing, hafnium disulfide, hafnium oxide, 2D layered material (2DLM), forming-free, compliance-free}\leavevmode\newline

\section*{Introduction}
The growing demand for data-driven computation in artificial intelligence (AI) and machine learning (ML) applications has skyrocketed energy consumption from modern data centres, with estimates showing that global yearly usage from this sector surpasses 200TW-hours per year. This figure is predicted to rise by an order of magnitude by 2030 \cite{mehonic2022masterplan} highlighting the growing need to address this problem \cite{jones2018stop}. One fundamental issue lies in the von Neumann architecture used for computation, which physically separates processing and memory units. The shuttling of data back and forth between such processing and memory units inevitably wastes energy and increases computation time. Neuromorphic computing aims to address this by taking inspiration from the human brain - a biological "computer" capable of performing complex tasks with far higher efficiency. For example, Google DeepMind's AlphaGo model required the use of 1,920 CPUs, 280 GPUs and consumes $\sim$ 1MW of power to play the boardgame Go (inference) against a human player. Comparatively, the human brain uses only $\sim$ 20W for the same task \cite{AlphaGOPowerConsumption} \cite{AlphaGoPowerConsumption2}. One feature of the brain directly responsible for this discrepancy is the co-location of processing and memory in an extensive network of neurons and synapses. In combination with other factors, this means that handling noisy and unstructured data is a relatively unremarkable task for the human brain, enabling us to quickly, reliably and efficiently tackle complex problems. Even when handled by the best artificial neural networks (ANN) running on modern hardware, most ML and AI implementations require extensive and costly training, consume large amounts of power and require frequent retraining \cite{ANNsNotGoodAt}. Therefore, it is clear that for computing tasks which benefit from quick and efficient processing of noisy input data, brain-like (neuromorphic) computation is necessary \cite{Phys4NeuroComp}.\\  

To implement neuromorphic computation, we can take inspiration from the brain and develop systems which allow computation in memory. A common approach is to use crossbar arrays of densely-packed non-volatile memory (NVM) devices in deep neural network (DNN) or spiking neural networks (SNN) \cite{NeuroCompUsingNVM}. Usually, the NVM devices forming the building blocks of crossbar arrays are memristors. Memristors are two-terminal, non-linear devices with inherent memory and represent a leading candidate for hardware implementations of neuromorphic computing \cite{adnanmemristorsreview}. They have the ability to modulate their resistance between distinct states by the application of an electrical stress, such as ramped or pulsed voltages. Crucially, they retain their programmed resistance even when the electrical stress is removed \cite{chua1971memristor}.\\ 

Non-volatile resistive switching in solid-state devices based on titanium oxide thin films has been observed as early as 1960 \cite{argall1968switching}, with the memristor first being formally described in 1971 by Leon Chua\cite{chua1971memristor}. More recent interest in ReRAM memristive technology was sparked in 2008 when devices based on thin titanium oxide films were once again investigated, this time directly addressing Chua's memristor framework \cite{strukov2008missing}, and later when memristors were integrated in silicon suboxide films \cite{adnanSiOxMemristors}. Since then, the field has flourished, with improvements on our understanding of the switching mechanisms of ReRAM devices \cite{zhang2021evolutionofHfO2Filament} \cite{analogHfO2memSchottky}, the performance of each individual element \cite{imperfection2DMemristors} \cite{robust2Dmemristors} \cite{MemristorsBasedon2DMs}, scaling \cite{WaferScaleMoTe2Memristors} \cite{zhu2023hybrid} and different materials systems. One particular material that we will investigate is hafnium oxide (HfO$_x$), sometimes referred to as hafnia. Hafnium oxide is part of a class of materials called high-k dielectrics. High-k dielectrics are so named as they posses a high dielectric constant compared to SiO$_2$, which allow using thicker layers in metal oxide semiconductor capacitors to suppress tunnelling while maintaining large capacitance \cite{HighkDielectrics}. In memristive applications HfO$_x$ - based devices fabricated by sputtering or atomic layer deposition (ALD) have already demonstrated low-energy switching, a high degree of scalability,  signal processing, image compression and convolutional filtering\cite{li2018analogue}, good endurance \cite{govoreanuLowEHfHfOx} and synaptic applications \cite{peng2021hfo2synapse} \cite{covi2016hfo2neuromorphic} although there has been a collective desire to increase programming linearity in synaptic applications and reduce the individual memristor power consumption \cite{liu2020hfo2}. One potential avenue of investigation for fabricating low-power memristors is to look at 2D layered materials.\\

Since the ground-breaking experiment where single layers of graphene were isolated and their electron transport investigated \cite{novoselov2004electric}, several more 2D layered materials (2DLMs) have been discovered with thousands more predicted to exist \cite{1000sOf2DMs}. All of these 2DLMs have unique properties which highlight a key interest in their research. 2DLMs present the opportunity to fabricate devices with atomic precision of thicknesses down to a single atomic layer, with chemically abrupt surfaces and no dangling bonds \cite{MemristorsBasedon2DMs} \cite{2DLMMemristorsReview} \cite{2DMsStateOfTheArt} \cite{liu2DMsNextGenComp} \cite{geimVDWHetero} \cite{purdie2018cleaning}. This allows devices with a wide range of properties and applications to be realised, combining different materials with varying properties with ease \cite{purdie2018cleaning}. Wafer-scale growth of desired 2DLMs for memristors by chemical vapour deposition (CVD) has been developed \cite{zhu2023hybrid}, with pilot lines already established. For example, memristors fabricated from wafer-scale CVD hexagonal boron nitride (hBN) have shown both high ON/OFF ratios of $10^{11}$ and energy consumption as low as 8.8zJ, demonstrating the potential for this technology \cite{2DLMMemristorsReview} \cite{zJWaferScaleCVDhBN}. However, sample-to-sample variation and poor control of defect densities present in the films makes them as-yet unsuitable for at-scale fabrication of memristors \cite{LanzahBNCVDQuality}. Another scalable fabrication technique that has been demonstrated with 2D materials of particular interest is the template growth of ultrathin oxides from semiconducting crystals. Starting from a near-perfect crystalline 2D layered material, native oxides can be formed by partially oxidising semiconducting crystals with almost perfect interfaces, using plasma oxidation \cite{AzizGaSPaper} \cite{wang2020atomicallyHfO2fromHfS2} \cite{liu2021HfSe2lowpower}, thermal oxidation \cite{jin2021controllingHfO2fromHfS2} or photo-oxidation \cite{PhotoOxLaserWrite}. Such devices have shown promising memristive behaviour. Recently, GaS$_x$O$_y$/GaS memristors with an ultrathin oxide and clean interface have been fabricated for the first time, showing low energy $\sim$ 0.22nJ operation \cite{AzizGaSMemristors}. Similarly, plasma oxidised HfO$_x$Se$_y$/HfSe$_2$ memristors have also been fabricated, demonstrating operation at low current compliance (100pA) and low energy (0.1fJ) in filamentary resistance switching \cite{liu2021HfSe2lowpower}.\\

Despite excellent performance being demonstrated in many memristors, devices which can be operated without the requirement for compliance or electroforming are even more desirable to reduce the power consumption, operation time and integration area cost of utilising memristors as synaptic weights in neuromorphic chips, improving efficiency \cite{SnOFormingCompFree}. The peripheral circuitry and transistor-memristor (1T1M) architectures required to integrate most ReRAM memristors into neuromorphic and ML chips provide a high barrier for their adoption \cite{ZnOformingfree} \cite{SnOFormingCompFree}. A few fabrication methods have shown promise in this regard. Methods which engineer a large proportion of oxygen vacancies in the pristine device such as H$_2$ annealing or using a Ti electrode gettering layer have been largely successful in promoting forming-free behaviour in ZnO \cite{ZnOformingfree} and Ta$_2$O$_5$ memristors\cite{ReRAMFormingFreeThin} respectively. However, the highly defective oxide switching layers inducing forming-free resistive switching also necessitate the application of a current compliance to prevent high currents from damaging the device \cite{ZnOformingfree} \cite{ReRAMFormingFreeThin}. SnO$_x$ ReRAM devices have been shown to exhibit both forming and compliance-free resistive switching. The forming-free behaviour is thought to be due to the ultra-thin oxide used \cite{SnOFormingCompFree} and the compliance-free nature of this device is thought to be a result of the TiO$_{2-x}$ interfacial layer acting as a series resistor, limiting currents and driving gradual resistive switching. However, many devices which exhibit these properties typically show a low R$_{ON}$/R$_{OFF}$ \cite{SnOFormingCompFree}, or employ fabrication techniques incompatible with CMOS \cite{Mos2analogpolycrystMemris} \cite{NonIdealON/OFFMemris}.\\

This presents a technological challenge: to fabricate devices which can combine strong non-volatile memory characteristics such as resistive switching at low energy and stable retention (found in many abrupt switching devices), while allowing for nearly continuous multi-level switching, high integration density and simplified operation without electroforming or compliance (which can be found in many gradual switching devices.) In this work, we address this by investigating the use of HfO$_x$S$_y$/HfS$_2$ structures where a thin HfO$_x$S$_y$ oxide is obtained by partial oxidation of 2D HfS$_2$ crystals for resistive memories. The oxidation of HfS$_2$ flakes has already been demonstrated by a wet oxidation (thermal oxidation in ambient) method \cite{wang2020atomicallyHfO2fromHfS2} and by a plasma partial oxidation method, demonstrating the application of such crossbar structures in flash memory\cite{jin2021controllingHfO2fromHfS2}. Instead, here we investigate a controlled dry oxidation method to produce HfO$_x$S$_y$/HfS$_2$ heterostructures to be used in memristive devices. We show that our devices behave as non-volatile memory devices without requiring an electroforming initialisation step or current compliance for their operation. The memristors show fast and low-energy operation, consuming as little as $\sim$20 pJ per switching cycle with 60ns programming pulses. Moreover, we also demonstrate a good R$_{ON}$/R$_{OFF}$ and stable state retention of multiple states over 10$^4$s at 150$^o$C. \\  

\section{Material Preparation and Characterisation}
% \subsection{Material Production and Characterisation}
HfS$_2$ flakes were deposited onto oxidized silicon wafers by mechanical exfoliation. The flakes were then oxidised using a low-temperature 110-minute dry thermal oxidation method, carefully controlling pressure, temperature, oxygen flow rate and time. The low temperature ($<300^oC$) oxidation process allows us to conduct this step even after transfer onto metal contacts and is well below the CMOS compatibility limit of $<450^o$C \cite{MaxTempCMOS}.\\

\begin{figure}[h!]
\begin{center}
\includegraphics[scale=0.8]{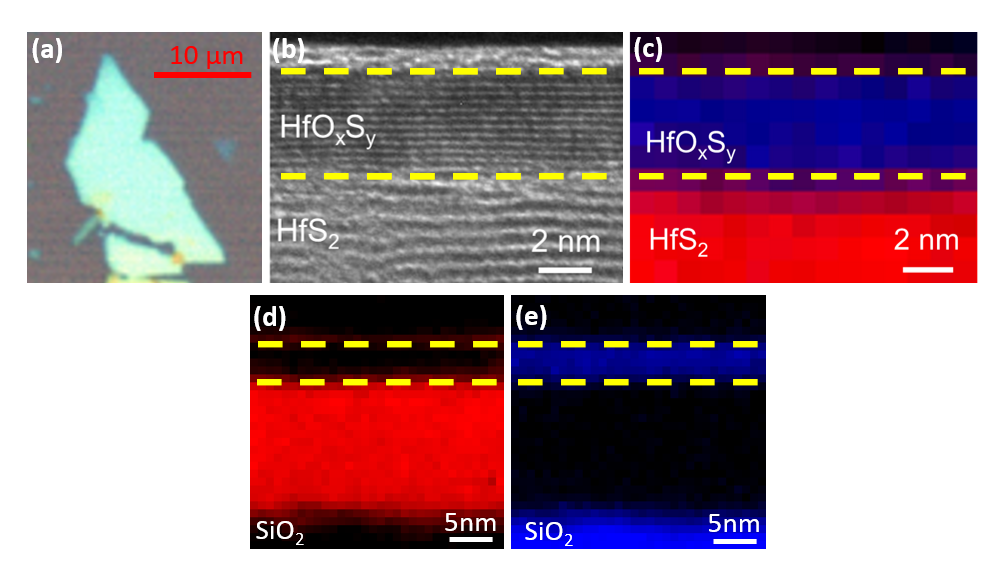}
\end{center}
\caption{(a) Optical microscopy image of HfO$_x$S$_y$/HfS$_2$ flake before being transferred to lamella for transmission electron microscopy (TEM). (b) Cross-sectional TEM image of partially oxidised flake shows crystallinity of resulting oxide layer and (c) electron energy-loss spectroscopy (EELS) mapping of (d) sulfur (red) and (e) oxygen (blue) demonstrates a clear separation of the HfO$_x$S$_y$ and HfS$_2$ regions.} 
\label{fig1}
\end{figure}

To gain insight into the structure of the oxide, we investigated a partially oxidised HfS$_2$ flake (Fig. \Ref{fig1}a) using cross-sectional transmission electron microscopy (TEM, Fig. \Ref{fig1}b). The atomic arrangement in the sample appears to be ordered in both the semiconductor and oxide regions, supporting the idea of a crystalline or large-domain polycrystalline oxide being formed, which is quite different from other oxides achieved by converting (predominantly by plasma and ambient/ wet thermal) 2DLMs to their native oxides \cite{wang2020atomicallyHfO2fromHfS2} \cite{xiongBPoxidAmorph} \cite{huhWSe2oxidAmorph} \cite{tianBPoxidAmorph} \cite{liu2021HfSe2lowpower} \cite{jin2021controllingHfO2fromHfS2}. The sample was oxidised under dry conditions but perhaps crucially oxidised at low temperature and cooled in an oxygen - free environment, as this has been seen to produce crystalline oxides in other transition metal dichalcogenides (TMD) oxidation-conversion work \cite{heWSe2oxidCrystal}. The oxide formed with the parameters detailed in the Methods is 4nm thick, however this is a process dependent property. The electron energy loss spectroscopy (EELS) maps (Fig. \Ref{fig1}c) show that there is a clear separation of semiconductor (Fig. \Ref{fig1}d) and oxide (Fig. \Ref{fig1}e) regions with a good interface, however in the top region of figures \Ref{fig1}c-d a sulfide layer can be observed. The origin of this region is not known, however it could be related to the displacement of sulphur species during oxidation. Furthermore, a small amount of sulphur is present in the oxide layer (Fig. \Ref{fig1}d), hence the designation HfO$_x$S$_y$. The process-dependent oxide thicknesses and characteristics were also studied with spectroscopic imaging ellipsometry (SIE) in good agreement with the TEM data (Supplementary Information, Fig. 1).\\

\section{Electrical Measurements}
\subsection{Pulsed Switching}
\begin{figure}[h!] 
\begin{center}    
\includegraphics[scale=0.66,trim=4 4 4 4,clip]{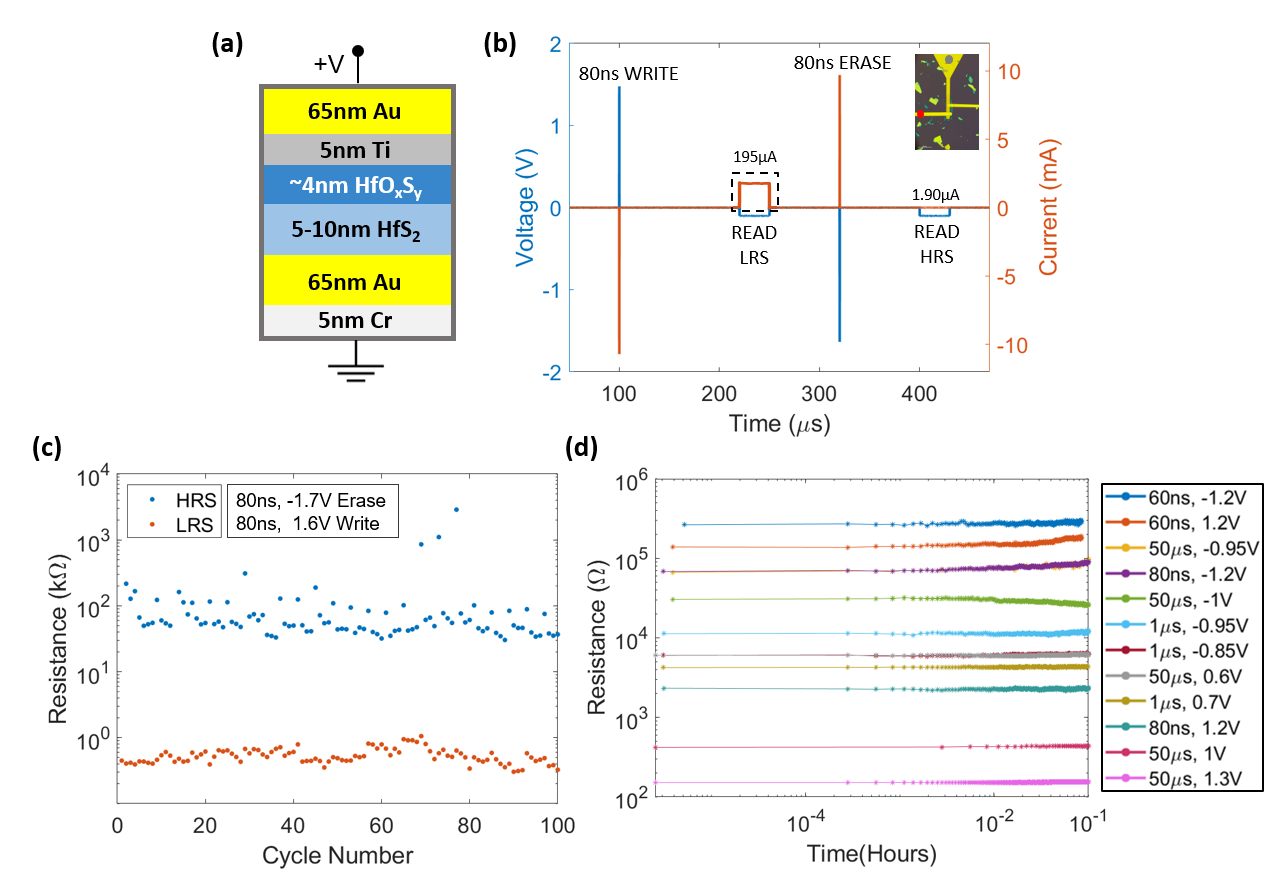}  
\end{center}
\caption{ (a) Schematic of typical devices fabricated using dry oxidation method. (b) Pulsed voltage test applied to the device with associated current response showing non-volatile WRITE and ERASE operations (READ currents expanded for visibility.) Energy=1.3nJ for WRITE and ERASE. (c) 100 cycle state distribution of LRS and HRS with 80ns, 1.6V WRITE pulses and 80ns, -1.7V ERASE pulses, resulting in an R$_{ON}$/R$_{OFF}$ of 102. (d) Multiple stable resistance states are achievable by varying pulse width and amplitude.}
\label{fig2}
\end{figure}

HfO$_x$S$_y$/HfS$_2$ memristive devices were fabricated using the oxidation method described previously (and in the Methods section), resulting in the structure shown in Fig. \ref{fig2}a. To examine the performance characteristics of our devices which determine their capability for integration in ML and neuromorphic applications, we applied voltage pulses and measured the current response in time. The voltage was applied to the top electrode while keeping the bottom electrode grounded. The device was SET from a high resistance state (HRS) to a low resistance state (LRS) during the WRITE operation, which was programmed as a pulse with a 20ns rise and fall time, held at 1.6V for 80ns (Fig. \Ref{fig2}b). The ERASE operation allowed the device to RESET its resistance from LRS back to HRS and was applied just as the WRITE operation, but at -1.7V instead. READ operations were performed at -0.1V over a significantly longer time (30$\mu$s) and all pulses were spaced $\sim$75$\mu$s apart to avoid any spurious contribution to the current originating from charge and discharge from parasitic capacitances. The device exhibited reproducible resistive switching over the 100 cycles tested (Fig. \Ref{fig2}c), switching between 0.51k$\Omega$ and 52.5k$\Omega$ resistance states for LRS and HRS respectively.\\ 

Importantly, no IV sweep characterisation or electroforming was conducted to initialise these devices prior to their initial pulsed testing to avoid any effect resulting from previous testing that might modify their switching behaviour. In fact, resistive switching has been achieved in these devices without electroforming or current compliance, which is unusual for HfO$_x$ - based devices and provides a significant advantage for their integration in circuits \cite{ZnOformingfree} \cite{SnOFormingCompFree} \cite{rajendran2023building}. Although many memristive devices show excellent performance and scalability, they require peripheral circuitry for electroforming, and integration into transistor-memristor (1T1M) architectures to limit high currents that would otherwise permanently damage the device. These factors not only hinder integration density and complicate peripheral control circuitry, but they also increase computation time and increase energy costs, reducing the applicability for many memristive devices into neuromorphic or dedicated ML chips \cite{SnOFormingCompFree} \cite{ZnOformingfree}. Dedicated circuitry for electroforming and current compliance would not be required to operate our thermally oxidised HfO$_x$S$_y$/HfS$_2$ memristors, highlighting one of the key merits of these devices.\\

By integrating the current over the pulse programmed WRITE and ERASE times and multiplying by the programmed voltage respectively, the WRITE and ERASE energies for each operation can be estimated to be $\sim$1.3nJ. A good R$_{ON}$/R$_{OFF}$ of 102 and resistive switching with minimal state conductance drift was observed (Fig. \Ref{fig2}c). This demonstrates that this device can be used for stable, low energy resistive switching with a good R$_{ON}$/R$_{OFF}$. 60ns pulse width WRITE and ERASE operations have also been investigated (see Supplementary Information, Fig. 3e). With 60ns switching, we observe even lower WRITE and ERASE energies of 16.5pJ and 17pJ respectively for the few cycles investigated. Such fast pulses are at the absolute maximum temporal resolution of the parameter analyzer and remote sensing unit setup (Keysight B1500A + waveform generator fast measurement unit - WGFMU) and for cross - point devices on Si$O_2$/ Si substrates which can exhibit significant capacitive effects. Therefore, all pulses widths employed were verified with an oscilloscope connected to a remote sensing unit on the voltage input terminal of the memristor (Supplementary Information Fig. 3a-d).\\

Fig. \Ref{fig2}d shows a variety of resistance states achievable by programming a single device with different pulse widths and voltages, and the retention characteristics of these states. Notably, the range of resistance states accessible in the device span 3 orders of magnitude, which is significant for circuit applications of multi-level memristive devices. Using this variable conductance behaviour in pulse-width or amplitude modulation pulsing schemes \cite{NonIdealON/OFFMemris}, further experiments should be conducted to demonstrate potentiation and depression with linear and high-granularity weight update, taking advantage of the wide range of stable resistance states available.\\

\subsection{Device Stability}
\begin{figure}[h!]
\begin{center}    
\includegraphics[scale=0.75,trim=4 0 0 0,clip]{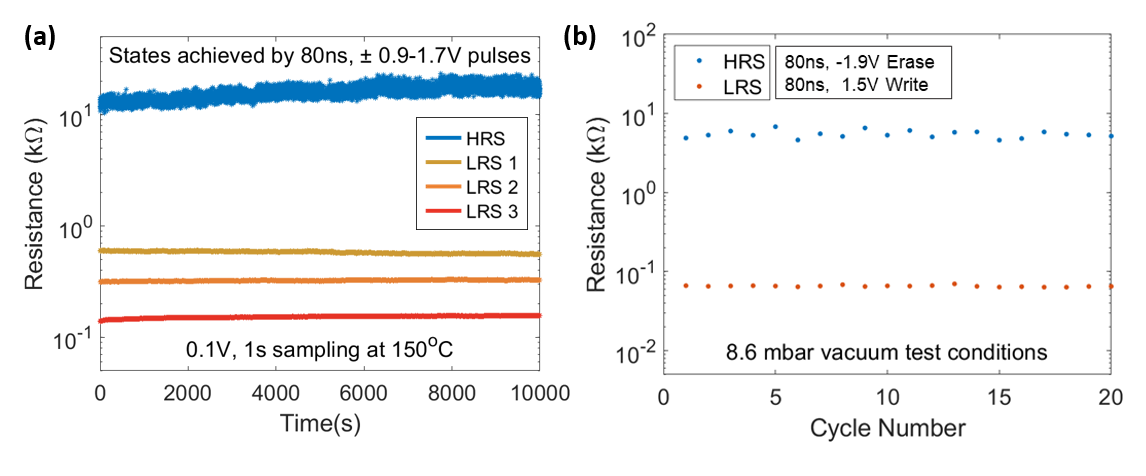}
\end{center}
\caption{(a) Retention of non-volatile states achieved by 80ns pulse programming, retained over 10$^4$s at 150$^o$C. (b) Pulsed voltage resistance switching unaffected by operating device in a 8.6 mbar vacuum environment.}
\label{fig3}
\end{figure}

For integration in modern machine learning hardware, a non-volatile memory device should not only show the ability to switch repeatedly between multiple distinct resistance states, but it should also show good retention of programmed resistance states and no dependence on external factors (other than applied voltage pulses) for its operation. To test the state retention of our devices, we programmed a few different resistance states using 80ns, $\sim\pm$0.9V - 1.7V pulses and observed the resistance drift at an elevated temperature in a moderate vacuum (Fig. \Ref{fig3}a). Resistance states were read at 0.1V every second over 10$^4$s. The LRS and HRS states at 150$^o$C observed in the device range from $\sim$20k$\Omega$ to the 0.15k$\Omega$ regime, respectively. Crucially, all the states investigated were stable at each temperature throughout the 10$^4$s testing period, despite the aggressive testing conditions. Testing the retention of the device at temperatures higher than room temperature accelerates the ageing processes inside the device which typically lead to failed state retention. Therefore, the stable retention at elevated temperatures demonstrates the potential for long retention of multiple resistance states in these devices.\\

To exclude the influence of water vapour or oxygen in the surrounding air from the device operation, we measured the devices in a 8.6 mbar vacuum environment (Fig. \Ref{fig3}b), applying similar programming pulses to before (80ns, 1.5V WRITE and -1.9V ERASE.) The operation of the device is unaffected by the vacuum environment, demonstrating that it is not dependent on any external factors such as oxygen or moisture. This is an important demonstration of feasibility for integration of HfO$_x$S$_y$/ HfS$_2$ devices with existing technological practices, such as the use of passivation layers in CMOS applications. Passivation layers bury the device in a dielectric material which is used to protect and stabilise them. Therefore, memristors integrated in neuromorphic chips or crossbar arrays would have to operate independent of their surroundings, in an oxygen-free environment, which we demonstrate here.\\

Overall, investigating devices by pulsed voltage is critical from a technological perspective: to demonstrate their capability for integration in ML and neuromorphic applications. Our HfO$_x$S$_y$/ HfS$_2$ memristors switch independetly of atmospheric oxygen and moisture, at low energy, responding to fast voltage pulses while providing a good R$_{ON}$/ R$_{OFF}$. They demonstrate tuneable conductance within a wide range of values and strong resilience to temperature and ageing over time. Crucially, we are able to operate these devices without an electroforming initialisation step or requiring any complex current compliance circuitry. This unique combination of desirable performance characteristics highlights the need for further investigation into wafer-scale and simulation-based implementations of such devices in neural networks for explicit machine learning or neuromorphic tasks. However, to design a scalable technological node which can harness these desirable characteristics, it is therefore equally important to understand the likely complex underlying mechanisms responsible for resistive switching in these devices.\\

\subsection{Voltage ramps}
\begin{figure}[h!]
\begin{center}    
\includegraphics[scale=0.75,trim=4 0 0 0,clip]{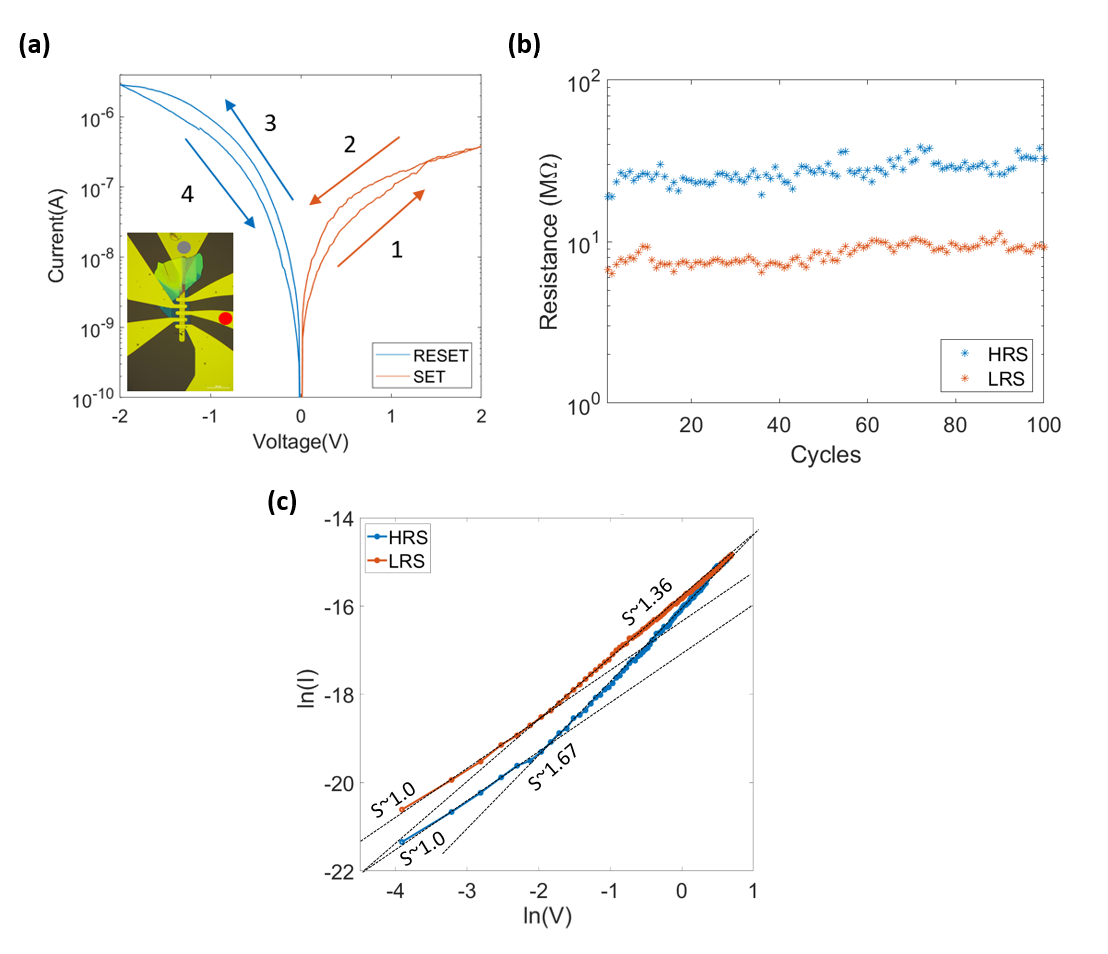}
\end{center}
\caption{(a) Forming and compliance - free, gradual resistance switching of the device shown in the inset. From (b), consistent resistive switching endurance can be observed, with the resistance gradually increasing over cycles. (c) The lack of an abrupt filament formation prevents a runaway region from appearing in the lnln IV plot of a representative SET cycle from which we can assume transport characteristics.}
\label{fig4}
\end{figure}

While pulsed voltage resistance switching demonstrates the performance and efficiency of the device in a more realistic application, the mechanisms behind the resistive switching in the device are better studied by ramping the voltage applied to the top contact of the device and inspecting the current output. This is particularly important for our devices as they employ an uncommon structure and morphology for a vertical memristive device. To do so, we tested a different device which was not previously exposed to any voltage stress. The device, shown in the inset of figure \Ref{fig4}a, is nominally identical to the one used for pulsed switching, with the only difference being the thickness of the semiconductor. The device was characterised using IV sweeps (Fig. \Ref{fig4}a) and the representative IV characteristics shown in Fig. \Ref{fig4}a were produced without requiring an initial electroforming step or a current compliance just as before and with the device initially in its HRS. The HRS and LRS evolution over 100 SET/RESET cycles are presented (Fig. \Ref{fig4}b).\\ 

The device operates at low SET and RESET currents, reading 26nA at -0.2V ($\sim$7.7M$\Omega$) in LRS compared to 7.4nA at -0.2V ($\sim$27M$\Omega$) in HRS (Fig. \Ref{fig4}b). The higher resistance of these devices compared to those in Fig. \Ref{fig2} is likely due to differences in device thickness and sample-to-sample variation. For the same oxidation parameters but higher initial HfS$_2$ thicknesses, the remaining thickness of un-oxidised HfS$_2$ in the device will be greater. This in turn, reduces peak currents in the device (Supplementary Information Fig. 6) presenting a method for tuning the conductance and energy consumption of the device.\\

The stark difference in the R$_{ON}$/R$_{OFF}$ in the more technologically relevant fast pulsed operation scheme (Fig. \ref{fig2} - higher R$_{ON}$/R$_{OFF}$) compared to the ramped voltage scheme used here to investigate transport in the device (Fig. \ref{fig4}) can be explained by the effect of Joule heating. Factors which are known to limit peak LRS conductance and hinder stability during switching such as Joule heating typically dominate over longer timescales \cite{hBNUltraFastRESETCurveBump}. This may explain the lower R$_{ON}$/R$_{OFF}$ of the device in Fig. \ref{fig4}, investigated with long-timescale IV sweeps, despite the higher voltages used in this instance. We have observed very similar R$_{ON}$/R$_{OFF}$ and IV sweep characteristics even in more conductive devices. Near the SET voltage of $\sim$1.45V in Fig. \ref{fig4}a, a small rise in conductance can be observed, but this is suppressed in IV sweep operation, likely contributed to by Joule heating and other conductance-limiting mechanisms in the device. Hence, maintaining fast switching timescales $\sim$80ns in pulsed operation allows for a larger conductance modulation than with voltage ramps (Fig. \Ref{fig4}) resulting in a higher R$_{ON}$/R$_{OFF}$ \cite{hBNUltraFastRESETCurveBump}.\\

In order to identify the dominating transport mechanisms, we analyse the natural logarithm of V and I for a representative SET voltage sweep, shown in Fig. \Ref{fig4}c. The characteristics of these IV sweeps are representative of the resistance switching observed in all the devices we have fabricated with this method. The HRS and LRS portions of the SET curve are represented by blue and orange lines respectively. Contrary to other similar devices \cite{liu2021HfSe2lowpower}, there is no abrupt and dramatic increase in conductance at the SET voltage, as described previously. This implies a conductance - mediated switching behaviour where, rather than breakdown effects such as electrode - bridging conductive filaments being formed, resistive switching is primarily caused by the overall field-induced redistribution of movable defects (such as oxygen vacancies) in the device \cite{MoS2analogSwitchingConductanceMed}. The electrical transport in the device seems to initially follow trap - controlled space charge limited conduction (SCLC), where there is an ohmic regime at low voltage such that ln(I) $\sim$ ln(V), then a region where the proportionality increases to $\sim$ 1.67 emerges - closer to the relationship described by Child's law. However after this step there is no dramatic runaway current region, yet the device still switches to a LRS. Most HfO$_x$ - based devices show Schottky emission in the high field HRS region \cite{liu2021HfSe2lowpower} \cite{hfo2zro2schottky}, however this is not the case in our devices. As opposed to the more commonly used (e.g. as gate dielectric) amorphous HfO$_x$ with low defect densities \cite{nodefectsinpristine} \cite{ZnOformingfree}, from TEM we have observed a crystalline oxide with sulfur defects present (Fig. \Ref{fig1}) and out-of-plane defect pathways (Supplementary Information, Fig. 2). Other 2D materials based memristive devices have shown the ability to produce low current and forming - free devices such as ours, due to pre - existing defect pathways along domain walls in the (poly)crystalline oxide \cite{Mos2analogpolycrystMemris}. Due to applied field, the conductivity of this pathway could be mediated by oxygen ion or vacancy diffusion \cite{li2022hafniaanalogmemris} as described in the conductance - mediated picture \cite{MoS2analogSwitchingConductanceMed}. As the applied positive voltage from the top electrode causes the oxide region to become more sub-stoichiometric due to oxygen ion migration into the Ti electrode \cite{valov2016mobile}, the resistance of the oxide region can be modulated without completely breaking it down or forming a continuous metallic filament \cite{MoS2analogSwitchingConductanceMed} \cite{li2022hafniaanalogmemris}.\\ 

In the LRS the electrical transport is also rather complex - showing an initial relationship of ln(I)$\sim$1.36ln(V) at high field, then an ohmic region at low field. This may be due to the semiconductor - oxide nature of our device, providing two additional junctions, one of which may be rectifying. These are not normally present in memristive devices, and the out-of-plane transport through the HfS$_2$ further complicates the picture.\\

\section{TCAD Simulation}
To further understand the transport and role of the semiconductor layer in our devices, we simulated resistive switching using a technology computer aided design (TCAD) software: Sentaurus (Synopsys, 2022). We compare a vertical memristor structure with a switching layer consisting of only HfO$_x$/HfO$_2$ to one more similar to our devices: consisting of HfS$_2$ as well.\\ 

The dynamics of three particles were simulated in our modelled devices under applied field in kinetic monte-carlo (KMC) simulations: mobile oxygen ions ($O^{2-}$), mobile oxygen vacancies ($V_O^{2+}$) and immobile, uncharged oxygen vacancies ($V_O$, composing the filament). Allowing for field - driven Frenkel pair ($O^{2-}$ ion and $V_O^{2+}$ oxygen vacancy) generation and recombination, we simulated resistive switching by ramping voltage in a quasistatic manner, dwelling at each voltage step for a certain period of time, recording the defect generation, particle movement in the device and the resulting conduction or leakage current (solving Poisson's equation) from top to bottom electrode. Spatially-dependent as opposed to spatially-agnostic vacancy generation was simulated by reducing the activation energy for Frenkel pair generation in the presence of pre-existing defects, guided by existing literature \cite{SchlugerHfO2FilaEasyActiv} \cite{zeumaultSentaurusHfO2Memristor}. Subsequently, filaments were generated as an energy-driven and spatially-dependent transition from free-moving, charged $V_O^{2+}$ particles into immobile, uncharged $V_O$ particles which can carry a conduction current. Trap-assisted tunnelling processes which would make the simulation more realistic \cite{zeumaultSentaurusHfO2Memristor} were omitted due to their heavy simulation time cost.\\

\begin{figure}[h!] 
\begin{center}    
\includegraphics[scale=0.8,trim=4 4 4 4,clip]{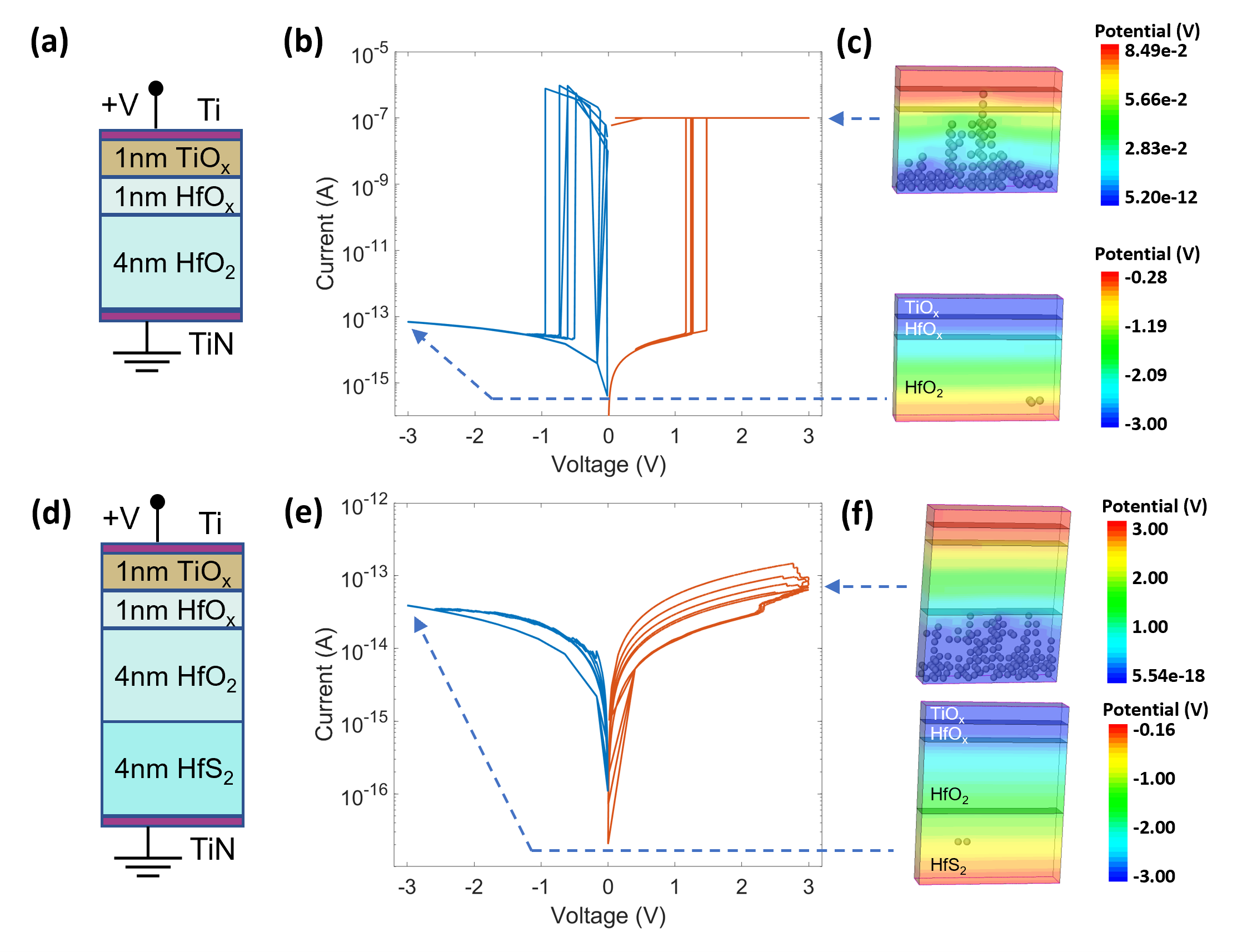}
\end{center}
\caption{Finite-element simulation of HfO$_2$ memristors without and with HfS$_2$ (a) Thicknesses of each layer in the HfO$_x$/HfO$_2$ memristors without HfS$_2$ (width = 1.5nm, length = 8nm). (b) Simulated IV sweeps of a device excluding HfS$_2$ exhibiting abrupt switching with filament formation and dissolution shown in (c) for peak SET and RESET voltages respectively. Electrostatic potential across device also shows modulation of potential due to the presence of the filament in device. Spheres in the electrostatic potential plots represent immobile vacancies ($V_O$) comprising the filaments. (d) Structure of simulated devices including HfS$_2$ region. (e) Simulated IV sweeps of a device using the semiconductor-oxide structure from (d) showing gradual resistive switching, with immobile vacancy concentration (represented by spheres) shown in the HfS$_2$ region in (f) for the peak SET and RESET voltages. Much smaller modulation of the potential due to vacancy aggregation in (f) can be seen compared to the electrode bridging filament in (c).} 
\label{fig5}
\end{figure}

We simulated device structures as shown in Fig. \Ref{fig5}a, using a design similar to existing HfO$_x$/HfO$_2$ memristors but with the lateral size being 8nm x 1.5nm to focus on the dynamics of a single filament and reduce simulation time. A Ti top electrode was modelled as an idealised contact of no thickness, on top of a 1nm thick Ti$O_x$ layer interfacing an equally thin HfO$_x$ layer. The latter two regions were included due to the well-known gettering effect of Ti electrodes on oxides \cite{stout1955getteringTiOxHfOx} \cite{valov2016mobile}. The Ti$O_x$ region therefore permitted in-and-out diffusion of $O^{2-}$ ions into the electrode under applied field to simulate this effect. Frenkel pair generation in HfO$_2$ based devices with Ti top electrodes has been shown to occur preferentially in the sub-stoichiometric HfO$_x$ region \cite{KMCSimFrenkelInterfaceHfO2TiOx}, so we simulate Frenkel pair generation only in the HfO$_x$ region, with the HfO$_2$ bulk allowing $V_O^{2+}$ diffusion and filament growth/ recession under applied field. The bottom contact was a TiN idealised contact.\\

In Fig. \Ref{fig5}b we show representative resistive switching cycles for a Ti/Ti$O_x$/HfO$_x$/HfO$_2$/TiN device. The device was cycled 5 times to demonstrate the general switching characteristics while saving simulation time. The device exhibits abrupt resistive switching and requires a current compliance to prevent complete dielectric breakdown, as is common in most HfO$_x$ - based memristive devices. The particles shown in Fig. \Ref{fig5}c are immobile vacancies ($V_O$) comprising the conductive filament, and the voltage was applied to the top electrode. Note that for both Fig. \Ref{fig5}c and \Ref{fig5}f, oxygen ions and mobile vacancies are not visualised for ease of interpretation.\\

With the inclusion of a HfS$_2$ region (Fig. \Ref{fig5}d), we observe compliance free, gradual resistive switching (Fig. \Ref{fig5}e) with tuneable device conductance (Supplementary Information, Fig. 6) as in our experimental devices. The HfS$_2$ region was simulated by starting with a template Si file and adjusting band gap (1.23 eV) and work function (5.71 eV) parameters due to limited available material parameters for few-layer HfS$_2$ \cite{FewLayerHfS2Transistors} \cite{2DMBandAlignmentHfS2} \cite{Type2HfS2Heterojunctions} \cite{BandAlignMonoTMDsHfS2}. Enabling charged vacancy diffusion and filament growth/ recession mechanisms in the HfS$_2$ region was critical for the replication of resistive switching similar to our experimental devices (Supplementary Information Fig. 4). A similar effect where charged oxygen vacancies migrated through HfO$_2$ and across the HfO$_2$/SiO$_2$ interface has been investigated \cite{ActivationEnergyVO2AndDifftoSiO2}. The vacancy was shown to stabilise in its neutral charge state upon entering the SiO$_2$ layer, just as we have modelled for the HfO$_2$/HfS$_2$ interface here. In our experimental devices, it is likely that vacancies may have migrated through out-of-plane defects present in the pristine device (Supplementary Information, Fig. 2) which could explain the forming-free behaviour.\\

Notably, simply increasing the thickness of the HfO$_2$ layer in Fig. \ref{fig5}a to match the HfO$_2$/HfS$_2$ thickness in Fig. \ref{fig5}d did not lead to any significant hysteretic behaviour in the sub-switching regime and the device still showed abrupt resistive switching otherwise. Therefore, the HfS$_2$ layer could play a vital role in limiting currents and preventing abrupt resistive switching, removing the need for current compliance. Greatly reduced peak current magnitudes can be observed in the HfO$_2$/HfS$_2$ device, however this has been shown to due to the small cross-sectional simulation area chosen (Supplementary Information Fig. 5.) By extrapolating to the lateral area of our experimental devices, we predict similar currents in the order of nA at 1.0 V. Furthermore we observe that the peak conductance of the device decreases with increasing HfS$_2$ thickness (Supplementary Information, Fig. 6), demonstrating the ability to tune the peak conductance and energy consumption of these devices based on their structure. The current observed is strongly related to the number of $V_O$ generated in the HfS$_2$ region in the simulation (Fig. \Ref{fig5}f). This suggests a conductance-mediated switching mechanism as the behaviour exhibited by the device is more akin to resistance  switching without complete filament formation \cite{li2022hafniaanalogmemris} \cite{analogHfO2memSchottky} \cite{MoS2analogSwitchingConductanceMed} rather than more abrupt and conventional filamentary bridging of the top and bottom electrodes through an amorphous hafnia switching layer. Video files showcasing the generation and diffusion of vacancies contributing to resistive switching in these devices can be accessed in the electronic supplementary material (Supplementary Movies 1 and 2, available in the UCL Research Data Repository, at https://doi.org/10.5522/04/27186993.v1) with a brief description of each movie in Supplementary Information section 5. The simulation accuracy and current magnitudes can be improved further by the inclusion of trap-assisted tunnelling processes such as elastic and inelastic trap to band and band to trap tunnelling of electrons, treating the $V_O^{2+}$ particles as electron traps which transition to $V_O$ particles upon electron capture \cite{zeumaultSentaurusHfO2Memristor}.\\

\section{Conclusions}
We have shown that oxide-semiconductor structures obtained by dry oxidation of HfS$_2$ at low temperature ($<300^{o}$C) can be used to realize forming-free and compliance-free resistive memories. By using voltage pulses, we demonstrate that non-volatile resistive switching with a good R$_{ON}$/ R$_{OFF}$ can be achieved with low energy, 80ns pulses. We also show that the devices can exhibit resistive switching with less than 20pJ pulses, highlighting their potential for low energy operation which is critical for ML and neuromorphic applications. Stable retention of resistance states at 150$^o$C has been observed, demonstrating the potential of these devices for long retention times and resilience to ageing. With the aid of IV sweeps and TCAD simulations, we investigate the transport in these devices and show that the semiconductor layer plays a key role in preventing oxide breakdown, enabling compliance-free operation and tuneable conductance/ power consumption. Further work is required to determine and decouple the exact switching and transport mechanisms in the device, however our work has shown that both the semiconductor-oxide structure and precise control of the oxide crystallinity play a significant impact on the performance and behaviour exhibited by 2D-materials based memristors, paving the way for simplified circuitry for neuromorphic computing.\\ 

\section{Methods}
\subsection{HfS$_2$ Production and Oxidation}
HfS$_2$ flakes were prepared by mechanical exfoliation of a HfS$_2$ crystal (HQGraphene) onto 285nm Si$O_2$ on Si in an inert ($N_2$) glovebox environment. The flakes were then examined by an optical microscope and transfer system within the glovebox. Oxidation was then performed inside a Moorfield Minilab 026 system also within the same inert environment glovebox. We performed dry oxidation at $280^oC$, 0.1mbar using 38SCCM $O_2$ for 110 minutes.\\

\subsection{Ellipsometry Measurements}
Ellipsometry measurements are performed with an EP4 Spectroscopic Imaging Ellipsometer (SIE, Park Systems Gmbh, Göttingen), equipped with a laser-stabilized Xenon arc lamp and a three-grating monochromator as the spectroscopic light source. This allows for an available spectral range between 360 nm to 1000 nm, and a $\sim$5 nm bandwidth of the output line. A polarization state generator (PSG), comprised of a linear polarizer and quarter-wave plate (compensator), controls the state of polarization of the incident beam. The reflected light is collected through an analyzer and a 50x objective to a 1392x1040 pixel CCD camera, allowing for a lateral resolution down to 1$\mu$m. All measurements are performed in the P-A-nulling mode, at an incidence angle of 50$^\circ$ with respect to the surface normal.\\

\subsection{TEM lamellar fabrication and characterisation}
The cross-sectional lamellas were prepared by a DualBeam FIB-SEM system (Carl Zeiss Auriga), equipped with a platinum (Pt) deposition cartridge and nanomanipulator (Kleindiek Nanotechnik). To minimize the ion beam damage, the sample was protected by a $\sim$100 nm platinum first by electron beam deposition (2 kV). After that, a several micrometer thick platinum layer was deposited using the gallium ion beam. After rough milling by 30 kV ion beam, the $\sim$2 $\mu$m thick plate was lifted out and attached to the edge of a copper finger, following thinning processes using a 15 kV ion beam to minimize the ion beam damage and the final lamella was less than 100 nm thick. After thinning, the lamella was transferred to the TEM column for observation as soon as possible to minimise potential further oxidation.\\

Transmission electron microscopy (TEM) and scanning TEM (STEM) were performed using uncorrected FEI Titan 80-300 with a Schottky field emission S-FEG source operated at 300 kV. Electron energy-loss spectroscopy (EELS) maps were carried out with a Gatan Quantum spectrometer with an energy dispersion of 0.5 eV per channel.\\

\subsection{Device fabrication}
Bottom electrodes were fabricated by a direct-write lithography method, evaporating 5nm of chrome as an adhesion layer, and then 65nm of gold by electron beam evaporation. Lift-off was performed in DMSO at $65^oC$ for 10 minutes. Flakes of HfS$_2$ used in devices were transferred by a polycarbonate (PC) stamp, which was melted onto bottom electrodes at 180$^o$C. The residual PC was dissolved in chloroform and the device cleaned by acetone and IPA cleaning. The flakes were heated in vacuum to a set point of 280$^o$C when oxidation commenced for 110 minutes at 0.1 mbar in a dry $O_2$ environment with flow rate of 38 SCCM in a Moorfield Minilab 026 chamber. The samples were then allowed to cool in vacuum. Finally, direct-write lithography was used to pattern the top electrode and 5nm of titanium then 65nm of gold was deposited by electron beam (e-beam) evaporation, with the lift-off performed the same as before.\\ 

\subsection{Device Characterisation}
Devices were measured on a FormFactor (formally Cascade) MPS150 probe station, connected to a Keysight B1500A Parameter Analyzer with remote sensing units (waveform generator/ fast measurement unit) with a temporal resolution of 10ns. Oscilloscope measurements were performed with a 1GHz Keysight p9243a oscilloscope. Measurements above room temperature were conducted with a 6-probe Nextron MPS-PTH chamber with Peltier stage.\\

\subsection{Sentaurus TCAD Simulations}
Simulations were performed in Synopsys Sentaurus TCAD. Device structures were designed in SDE, with a mesh size of 0.5nm and no doping. A template Silicon parameter file was used to represent HfS$_2$ in the device structure. Ti and TiN electrodes were simulated as idealised contacts. In SDevice, Kinetic Monte Carlo (KMC) simulations were used to show the time evolution of oxygen vacancy and oxygen ion generation, diffusion, and filament growth/ recession dynamics dependent on applied voltage to the simulated structures. SVisual TCL command files were used to produce particle plots, IV characteristics and animations are included as electronic supplementary material in the UCL Research Data Repository, at https://doi.org/10.5522/04/27186993.v1.\\ 

\section*{Competing interests} 
The authors declare that they have no competing interests.\\

\section*{Data availability}
Data for this article, including scripts used to plot figures in the main text and supporting information are available in the UCL Research Data Repository, at https://doi.org/10.5522/04/27186993.v1.\\

\section*{Acknowledgment}
We acknowledge helpful discussions with Abin Varghese and Bipin Rajendran. We acknowledge funding from EP/T517793/1. VN and XG wish to thank the support of the Science Foundation Ireland funded AMBER research centre (Grant No. 12/RC/$2278_P2$), and the Frontiers for the Future award (Grant No. 20/FFP-A/8950). Furthermore, VN and XG wish to thank the Advanced Microscopy Laboratory in CRANN for the provision of their facilities.\\

\bibliography{Main.bib}
\bibliographystyle{unsrt}

\end{document}

\maketitle
\section{Spectroscopic Imaging Ellipsometry}
\begin{figure}[h!]
\begin{center}
\includegraphics[scale=0.65]{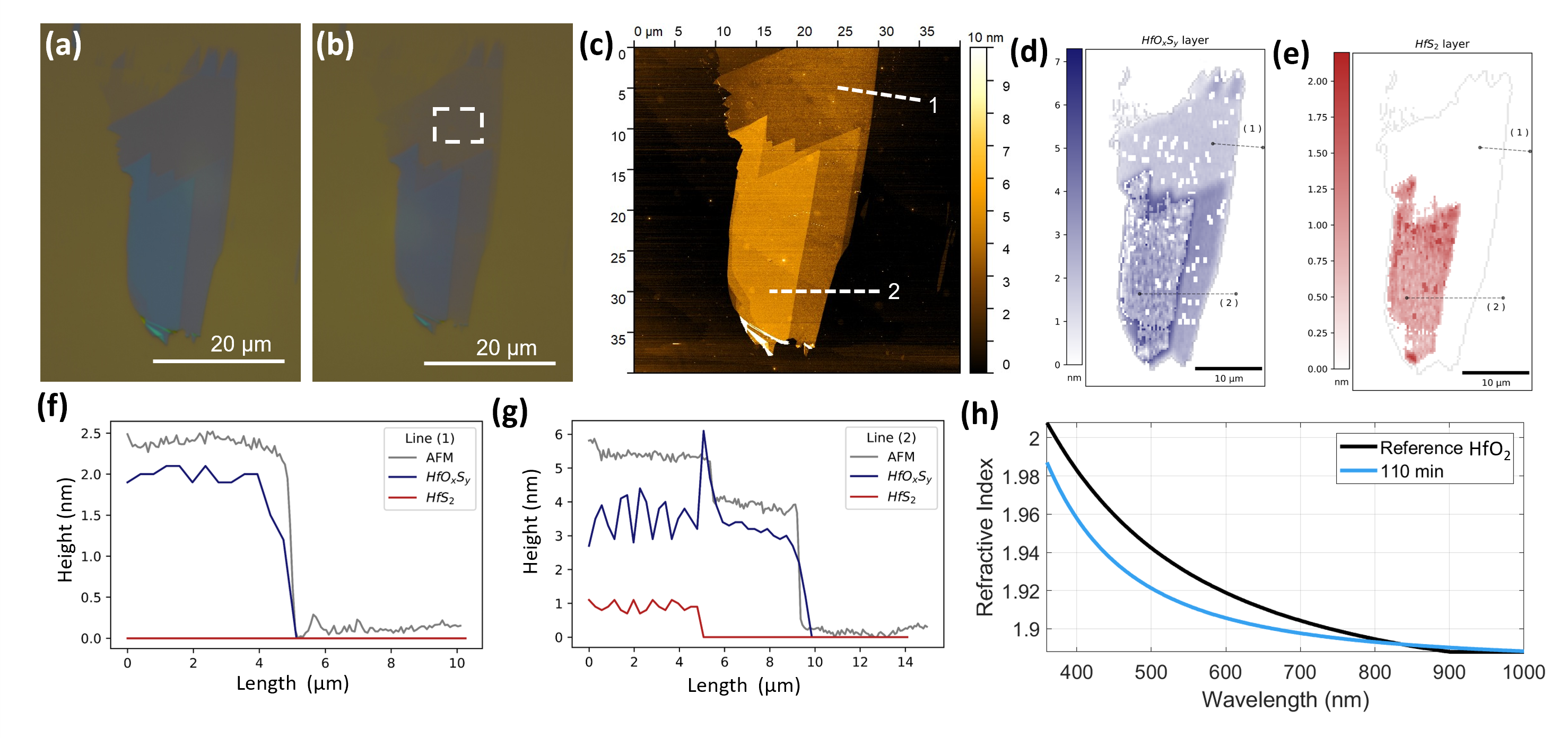}
\end{center}
\caption{(a) HfS$_2$ flake prior to oxidation shows a blue colour, with opacity dependent on layer thickness of the material. (b) Post 110 min dry oxidation, oxidised material appears semi-transparent. The slightly brighter circle at the centre of the images is a microscope artifact and not a feature of the material. The boxed region indicates the region where the surface roughness was calculated by AFM (c). (d) SIE confirms that oxidation is uniform across the flake within the resolution of the instrument and that (e) thicker regions retain HfS$_2$. (f-g) AFM and SIE thickness profiles of the HfO$_x$S$_y$ and HfS$_2$ linescans from figures c-e show that thicker regions retain HfS$_2$, whereas thinner regions are fully oxidised  with an oxidation depth of $\sim$4nm. (h) Refractive index comparison between the 110 minute dry-oxidised flakes and a reference HfO$_2$ sample shows that a good agreement of the optical properties to a reference value can be achieved.}
\label{Supfig1}
\end{figure}

Figures \Ref{Supfig1}a-b show the characteristic increase in transparency of a HfS$_2$ flake due to thermal oxidation. 
AFM images (c) were taken to determine flake thickness and surface roughness after oxidation (SI). A surface roughness of 125pm was determined in line with other work on the oxidation of HfS$_2$ \cite{wang2020atomicallyHfO2fromHfS2} \cite{jin2021controllingHfO2fromHfS2}.\\  

Spectroscopic imaging ellipsometry (SIE) was employed for the high-resolution imaging and characterisation of partially and fully oxidised HfS$_{2}$ samples, measured in ambient conditions at room temperature. Each map pixel encodes a complete set of ($\Psi$, $\Delta$) ellipsometric angles, which was fitted to a HfO$_x$S$_y$/HfS$_2$/SiO$_2$/Si optical model \cite{IrinaEllipsometry} using the EP4Model software. The thickness information was extracted for each layer as 3D maps (Fig. \Ref{Supfig1}d, e). The HfO$_x$S$_y$ map shown in Fig. \Ref{Supfig1}d confirms uniform oxidation across the entirety of the flake within the instrument resolution. No binning (defined as integrating N\,x\,N pixels of the original image into one pixel of the processed image) was applied during the data acquisition or image processing steps in order to maximise resolution, resulting in regions with signal loss (white pixels within the flake boundary). The complementary HfS$_2$ map from Fig. \Ref{Supfig1}e demonstrates thicker regions still have some remaining HfS$_2$ signal, implying an oxide penetration depth of $\sim$4nm. In figures \Ref{Supfig1}f-g we show the linescans 1 and 2 from figures \Ref{Supfig1}c-e. The data contained shows the height measured by AFM and the SIE-derived HfO$_x$S$_y$ and HfS$_2$ height profiles. At every point, the total sum of HfO$_x$S$_y$ and HfS$_2$ thickness matches almost perfectly with the measured thickness of the flake estimated from AFM scans (Fig. \Ref{Supfig1}c). The observed discrepancies in measured flake height from AFM and ellipsometry measured thicknesses (Fig. \Ref{Supfig1}f-g) could originate from measurement noise, surface roughness and contamination, or discrepancy between the modelled oxide properties and the measured sample. Between 5-9$\mu$m in Fig. \Ref{Supfig1}g, we see that for regions of the flake less than or equal to $\sim$4nm thick, the flake is completely oxidised as no sulfide is present in the ellipsometry signal. For thicker regions, such as the region between 0-5$\mu$m in Fig. \Ref{Supfig1}g, we see the remainder of the flake height is filled by unoxidised HfS$_2$. The increased oxide thickness observed along the flake edge is disregarded as a common SIE measurement artefact, as supported by AFM.\\

The complex refractive indices of the as-grown HfO$_x$S$_y$ samples were extracted from the measured ellipsometric spectra, and were fitted using the Cauchy dispersion model, assuming a transparent dielectric layer in the probing wavelength range. The results were compared to the optical properties of HfO$_2$ as given on the publicly available SOPRA database \cite{EllipsDatabase} (Fig. \Ref{Supfig1}h), showing a close agreement for the 110 minute dry oxidation recipe used in our devices.\\ 

\section{Out-of-plane defects}
\begin{figure}[h!] 
\begin{center}    
\includegraphics[scale=0.4]{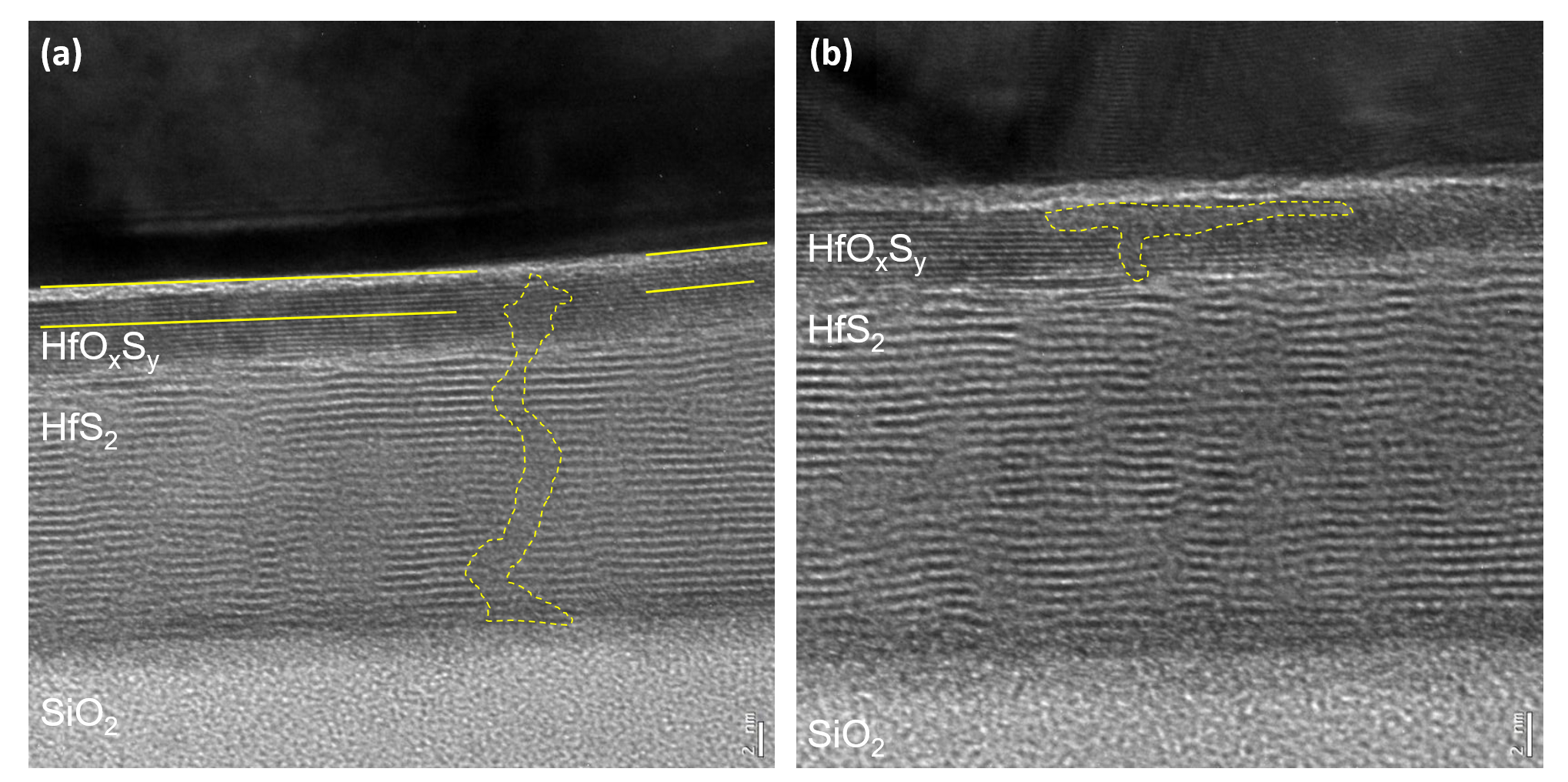}
\end{center}
\caption{Out-of-plane defect paths in 110-minute oxidised Hf$O_xS_y$/ Hf$S_2$ structure can occur (a) throughout the flake and (b) within the oxide.}
\label{Supfig2}
\end{figure}

We have observed out of plane defect paths in our 2D layered materials used for memristive devices described in the paper (Fig. \Ref{Supfig2}). In-plane defect paths such as grain boundaries bridging electrodes have been observed by Sangwan et al \cite{sangwanMoS2PolyCrysGrainBoundaryMemris}. They found that vacancies migrating withing the grain boundary could result in resistive switching due to the aggregation of charged defects under applied field. Similar to other devices \cite{analogHfO2memSchottky} \cite{MoS2analogSwitchingConductanceMed}, this would lead to a modification of the Schottky barrier height (SBH) \cite{sangwanMoS2PolyCrysGrainBoundaryMemris} \cite{IntrinsicMemMechGrainBoundary2DMaterialsReview}, showing very similar IV sweep behaviour to our devices, although we do not observe Schottky emission as a transport mechanism in the HRS or LRS. Out-of-plane defect paths have been shown to enable memristive switching by the same vacancy migration mechanism and have also been observed in multi-layered 2D layered materials such as hBN \cite{MultiLayerhBNGrainBoundMemris} \cite{OtherMultiLayerhBNGrainBoundMemris}.\\
  
\section{Pulsed Measurements}
\begin{figure}[h!] 
\begin{center}    
\includegraphics[scale=0.6,trim=0 0 4 0,clip]{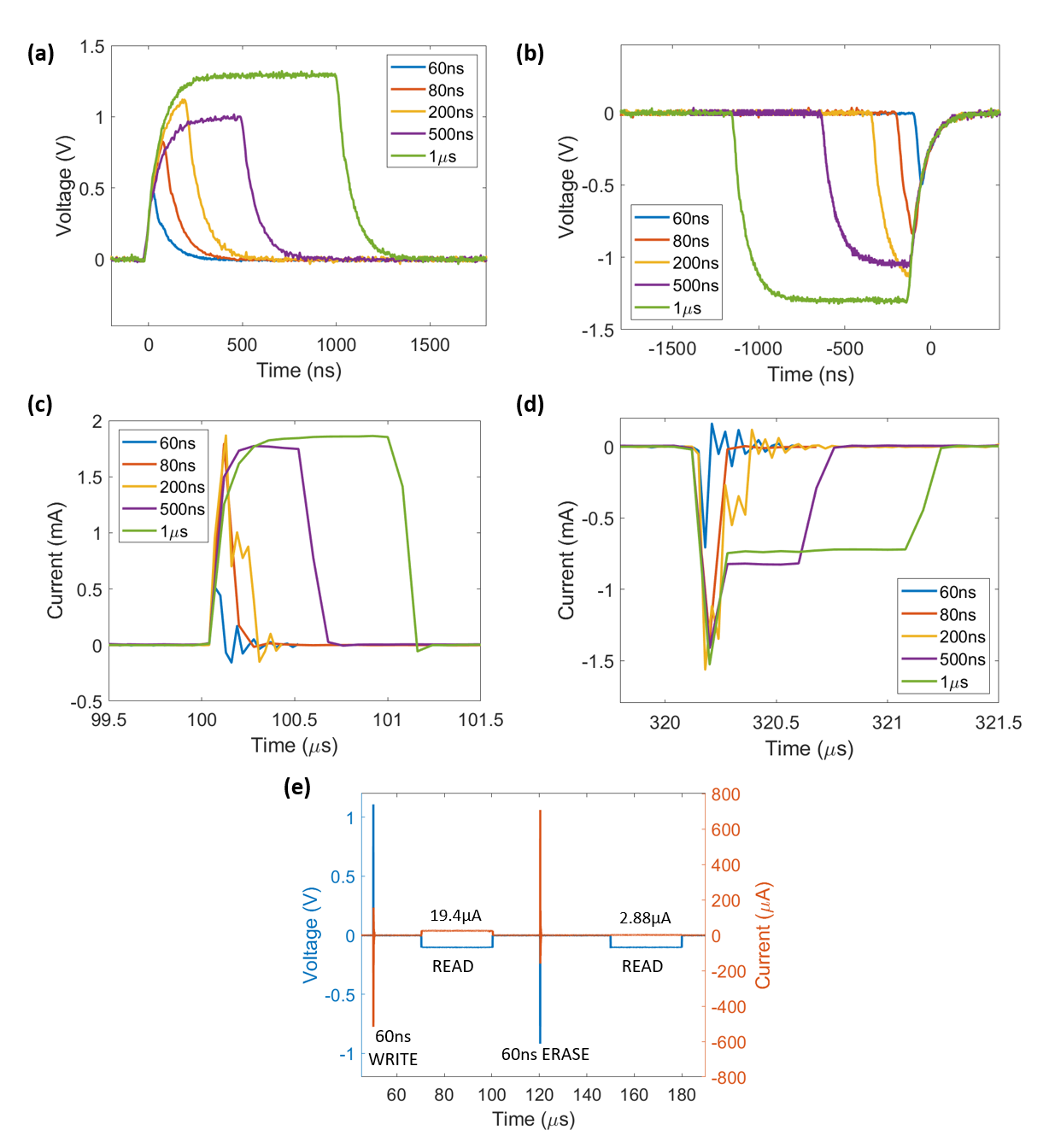}
\end{center}
\caption{With the use of an oscilloscope, WRITE (a) and ERASE (b) input voltage waveforms used to switch devices were investigated in high time resolution. By looking at current vs time on a parameter analyzer, we can see that a significant portion of the resistive switching occurs in the first 80ns for both WRITE (c) and ERASE (d). Resistive switching has been demonstrated down to 60ns (e) with comparable characteristics to other pulsing schemes, but lower energy.}
\label{Supfig3}
\end{figure}

The pulse times for different waveforms generated by the waveform generator fast-measurement unit (WGFMU) integrated in the Keysight B1500 parameter analyzer with remote sensing units were examined under a 1GHz Keysight p9243a oscilloscope (Fig. \Ref{Supfig3}a,b). The oscilloscope was connected to a remote sensing unit on the voltage input terminal of the memristor.\\

Generally, the pulse times were true to the pulse program sent from the WGFMU in total pulse width, however the pulse shapes were not square as they were programmed. WRITE and ERASE characteristics were examined (Fig. \ref{Supfig3}c,d) and we found that - much like other papers on fast-switching 2D materials-based memristors \cite{hBNUltraFastRESETCurveBump}, the RESET occurs within the first part of the ERASE pulse. The latter part likely predominantly contributes to Joule heating, and is exhibited in all pulses that are long enough to contain the full extent of the $\sim$80ns RESET peak seen in the first portion of the WRITE and ERASE pulses. Finally we show a representative pulsed switching cycle obtained by 60ns, $\sim\pm$1V pulses (Fig. \ref{Supfig3}e) on the devices showcased in the main text. The characteristics shown are largely similar between all the fast pulsed switching programs investigated. Recorded currents and energy consumption from 60ns switching (within the 80ns RESET threshold) are even lower than the other pulsing schemes investigated. Therefore, shorter timescales with dedicated sub-nanosecond pulsed switching units should be investigated further.\\

\section{TCAD Simulations}
A suite of simulations with sequential modifications made to the memristor materials stack to elucidate the role of each layer used in our Sentaurus TCAD simulations (Synopsys, 2022).
\begin{figure}[!htb] 
\begin{center}    
\includegraphics[scale=0.6,trim=0 0 4 0,clip]{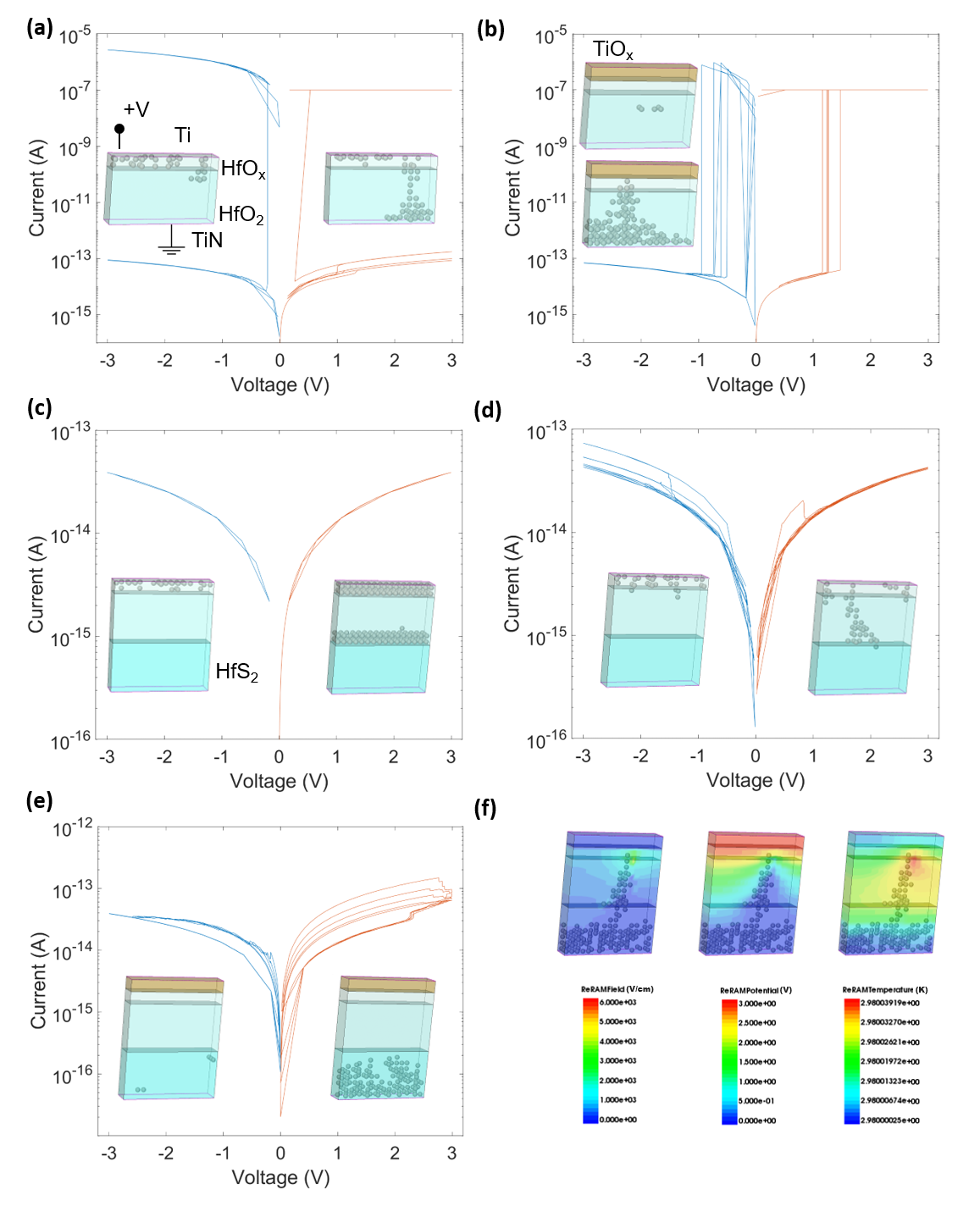}
\end{center}
\caption{(a) Without the Ti$O_x$ region, only breakdown occurs in the Hf$O_x$/ Hf$O_2$ device. (b) With Ti$O_x$, abrupt resistive switching is observed. (c) Without Ti$O_x$ but with Hf$S_2$, we see only leakage current with no hysteresis (note, particles shown only in this subfigure are mobile vacancies in Hf$O_x$ during RESET, then oxygen ions in Hf$O_x$ and vacancies on the Hf$O_2$/ Hf$S_2$ interface during SET. (d) Allowing filament growth and recession in Hf$S_2$ enables hysteretic behaviour but with poor filament stability. (e) Gradual and consistent resistive switching can be observed when adding Ti$O_x$, replicating the structure of our devices. (f) Electric field, electrostatic potential and temperature variation in device as a result of filament growth.}
\label{Supfig4}
\end{figure}\\

Omitting the Ti$O_x$ region (Fig. \Ref{Supfig4}a) provides no method for creating an imbalance between oxygen vacancies and ions in the switching layer of the device. This results in dielectric breakdown and vacancy filament formation, but prevents resistive switching using this biasing scheme. Including the Ti$O_x$ region thus enables stable and repeatable resistive switching (Fig. \Ref{Supfig4}b). By sequentially adding Hf$S_2$ and allowing vacancy diffusion throughout the device (Fig. \Ref{Supfig4}c), enabling stabilisation of mobile vacancies by transitioning to immobile vacancies/ filament formation (Fig. \Ref{Supfig4}d) and finally adding Ti$O_x$ (Fig. \Ref{Supfig4}e), the IV characteristics of our experimental devices can be replicated. Electric field, electrostatic potential and temperature variation in our device during filament growth are shown in a scenario where filament formation occurs almost from bottom to top electrode (Fig. \Ref{Supfig4}f). However they represent an area for improvement as the reported n-type conductivity of the conductive filament in hafnia as reported in other work \cite{zeumaultSentaurusHfO2Memristor} has not been replicated.  \\

\subsection{Area Scaling}
\begin{figure}[h!] 
\begin{center}    
\includegraphics[scale=0.8]{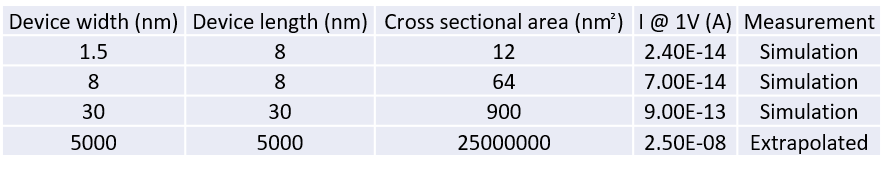}
\end{center}
\caption{Extrapolated from 3 simulations of the same device with different cross sectional areas, we can see that the current through the device at 1V in simulation is more similar to our experimental devices. This presents a possibility for reducing currents (and therefore energy consumption) in the device by scaling electrode size down.}
\label{Supfig5}
\end{figure}

We show that by extrapolation of 3 data points for the same device as Fig. \Ref{Supfig4}e, when scaling by area, a similar current exhibited by our experimental devices can be achieved (main text, Fig. 4a) Further data points were not taken as increasing the device cross sectional area beyond 8x8$nm^2$ presented a significant cost to simulation time (Fig. \ref{Supfig5}.) \\
\\

\subsection{HfS$_2$ Thickness Scaling}
\begin{figure}[h!] 
\begin{center}    
\includegraphics[scale=0.8]{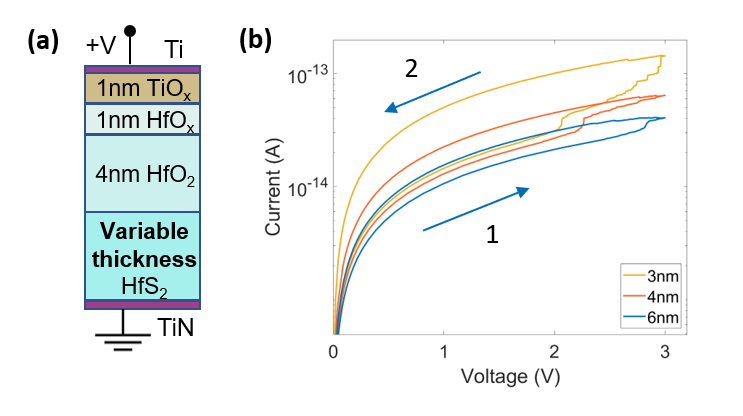}
\end{center}
\caption{(a) Device structure with variable HfS$_2$ thickness. (b) Representative IV SET sweeps from otherwise identical devices with 3nm, 4nm and 6nm thick HfS$_2$ layers, showing decreased conductance and peak currents for devices with thicker HfS$_2$.}
\label{Supfig6}
\end{figure}

By investigating devices identical to Fig. \Ref{Supfig4}e but with a HfS$_2$ region of variable thickness (Fig. \Ref{Supfig6}a), we show that an increase of 3nm in HfS$_2$ thickness can decrease the peak conductance of the device by 3.5X (Fig. \Ref{Supfig6}b).\\

This has been observed in our experimental devices where HfS$_2$ flakes of different thicknesses have been used to fabricate otherwise identical devices, supporting the idea of the HfS$_2$ region acting to limit currents and drive the gradual, compliance-free and forming-free resistive switching we have observed.\\

\section{Description of Additional Supplementary Files}
Supplementary Movies 1 and 2 detailed in this section are available as electronic supplementary material in the UCL Research Data Repository, at https://doi.org/10.5522/04/27186993.v1.\\
\subsection{Supplementary Movie 1}
Synchronised IV characteristics of a Ti/TiO$_x$/HfO$_x$/HfO$_2$/TiN memristor showing abrupt, filamentary resistive switching due to charged oxygen vacancy generation and diffusion under applied field (spheres in the device structure). The charged, mobile oxygen vacancies transition to uncharged, immobile oxygen vacancies and form a filament bridging the top and bottom electrodes, creating a conductive pathway. A compliance is applied in the simulation to prevent complete dielectric breakdown. Bipolar resistive switching is demonstrated as oxygen ions diffuse and recombine with oxygen vacancies (not visualised for clarity) under applied field of opposite polarity.\\
\subsection{Supplementary Movie 2}
Synchronised IV characteristics of a Ti/TiO$_x$/HfO$_x$/HfO$_2$/HfS$_2$/TiN memristor showing gradual resistive switching due to mobile, charged oxygen vacancy generation and diffusion under applied field (spheres in the device structure). The charged, mobile oxygen vacancies accumulate in the HfS$_2$ region resulting in a gradual modulation of the conductance of the device, but they do not form an electrode-bridging filament as in Supplementary Movie 1. Bipolar resistive switching is demonstrated as oxygen ions diffuse and recombine with oxygen vacancies (not visualised for clarity) under applied field of opposite polarity.\\

\bibliography{SI}
\bibliographystyle{unsrt}